# Modulation above pump-beam energy in photoreflectance

D. Fuertes Marrón [1]

*Insituto de Energía Solar, Universidad Politécnica de Madrid, ETSIT, Ciudad Universitaria s/n, Madrid, 28040, Spain*

**Abstract.** Photoreflectance is used for the characterisation of semiconductor samples, usually by sweeping the monochromatized probe beam within the energy range comprised between the highest value set up by the pump beam and the lowest absorption threshold of the sample. There is, however, no fundamental upper limit for the probe beam other than the limited spectral content of the source and the responsivity of the detector. As long as the modulation mechanism behind photoreflectance does affect the complete electronic structure of the material under study, sweeping the probe beam towards higher energies from that of the pump source is equally effective in order to probe high energy critical points. This fact, up to now largely overseen, is shown experimentally in this work. $E_1$ and $E_0+\Delta_0$ critical points of bulk GaAs are unambiguously resolved using pump light of lower energy. This type of *upstream* modulation may widen further applications of the technique.

Photoreflectance (PR) is a pump & probe spectroscopy well known in the characterization of semiconductor materials and devices [1,2]. It relies on the diffusion of charge carriers photogenerated with the pump beam and the subsequent screening of electric fields already present in the sample at space-charge regions, typically located at interfaces and free surfaces. The dielectric constant of the specimen, and thus its reflectance $R$, are slightly perturbed upon the field modulation. Such small changes in reflectance, $\Delta R$, are detected using phase-sensitive techniques with a probe light beam swept in wavelength and typically expressed as relative $\Delta R/R$ ratios. The technique contributed significantly to the present understanding of the electronic structure of most typical semiconductors [3] and has found continuity as a valuable characterisation tool of novel materials, like dilute nitrides [4], low-dimensional structures [5,6] and their potential applications [7]. The detection stage in PR largely relies on the rejection of any pump light scattered upon interaction with the sample that may eventually end up at the detector. Scattered pump light is typically the main source of background noise, together with sample luminescence, in the resulting spectra [8], as it enters right at the chopping frequency tracked by phase sensitive detection. The use of long pass filters (LPF) right in front of the detector is commonplace in order to avoid such spurious scattering. PR proceeds thereof by sweeping the monochromatized probe beam toward lower energies from the uppermost value set by the filter edge, recording changes in reflectance of the probe upon the action of the pump beam. Implicitly, the highest energy accessible to the experiment is therefore set by the optical edge of the LPF, normally chosen a few hundreds of meV below the nominal photon energy of the pump source. This small offset accounts for both the line broadening of the source (particularly if LEDs are used) as well as the finite width of the filter optical edge.

---

[1] Author to whom correspondence should be addressed. Electronic mail: dfuertes@ies.upm.es.

In contrast to this sort of standard PR, the so called "first derivative" modulation spectroscopies [9], like piezo- (PzR) or thermo-reflectance (TR), do not appear bounded at high energies as a result of the perturbing action. In the case of piezoreflectance [10], stress-strain cycles are imposed on the sample, usually by means of a piezoelectric actuator attached to the sample, whereas in thermoreflectance [11] the sample is subject to thermal cycles induced, e.g., by a Peltier element. The same applies to electroreflectance (ER), making use of an externally applied modulated electric field on the sample [12]. Even when each modulation mechanism is executed at a reference frequency thereby used for detection, the detection itself is in principle not constrained to a certain photon energy range of the probe beam. The only practical limitations are imposed by the spectral content of the source and the responsivity of the detector employed. The reason is that the perturbation used as modulation agent, independently of its origin, does affect the entire electronic structure of the sample under test. PR is not different from PzR, TR or ER in that respect. The generation of photovoltage upon pump illumination of a semiconductor, on which PR is based, is better illustrated as a change in band bending at those regions in the sample sustaining space charge (SCR), typically free surfaces or interfaces, as schematically shown in Figure 1. Even when photogeneration of free carriers upon appropriate illumination may just involve the first interband transitions allowed between occupied and empty states, the entire electronic structure of the material is thereby affected, as long as the modulation of the electric field associated to the SCR is active. It is thus expected that electronic transitions at energies higher than those directly accessible with the pump beam be equally subject to the modulating action and consequently not PR-silent, as schematically shown in Figure 1. In other words: *upstream* modulation using probe photon energies higher than that of the pump beam should be equally accessible as in *downstream* PR using LPF, should the photon energy of the pump beam be sufficient in order to develop a measurable photovoltage. The latter can actually happen at the fundamental absorption edge of the sample or via defect states at sub-bandgap energies. In what follows we show evidence of the modulation of high energy critical points showing up in PR spectra of GaAs when using pump light of lower energy.

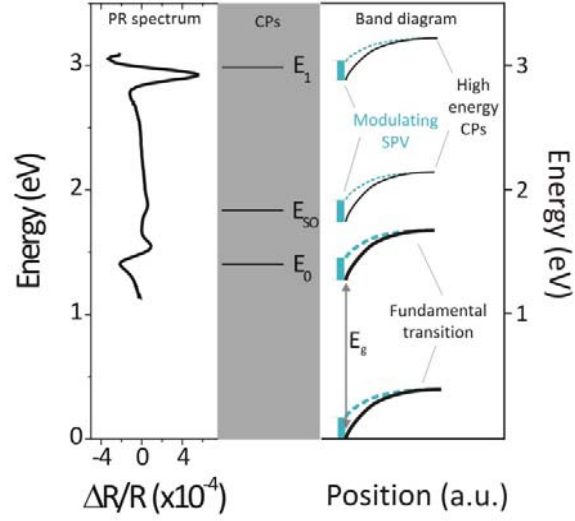

FIG. 1 Schematic representation of critical points (CPs) $E_0$, $E_0+\Delta_0$ (indicated as $E_{SO}$), and $E_1$ of GaAs represented in ascending energy on the same energy axis of a typical experimental PR spectrum (left). The band diagram picture (right) illustrates the modulation mechanism at CPs, namely periodic SPV generation, upon illumination with a chopped pump beam of energy slightly above $E_g$, inducing transitions at the fundamental gap. Notice that, although the $E_0+\Delta_0$ transition involves a valence state below the valence band edge (reference at zero energy), the corresponding energy is greater than the fundamental gap $E_g$.

For this purpose, we have used a Si-doped GaAs wafer (AXT, $n=1\times10^{18}$ cm$^{-3}$). The reason is that *n*-type doped GaAs exhibits intense and broad signatures in PR at room temperature, particularly in the range of $E_1$ transitions, that are typically better resolved than in intrinsic material. PR was measured using the light beam of a quartz-tungsten-halogen lamp (operated at 150 W) as probe of intensity $I_0(\lambda)$. The light is passed through a monochromator (1/8 m Cornerstone-Newport) and focused with optical lenses on the sample. Light directly reflected with intensity $I_0(\lambda)R(\lambda)$ is focused on a solid-state Si-detector. The current signal is transformed into a dc-voltage and preamplified (Keithley). The pump beam from a laser source is mechanically chopped at 777 Hz and superimposed onto the light spot of the probe on the sample, providing the periodic modulation. Three laser sources have been used as pump in the experiments, the 325 nm line of a 15 mW He-Cd laser, the line at 632.8 nm of a 30 mW He-Ne laser, and a solid-state laser diode operating at 814 nm. The signal recorded at the detector contains therefore two components: the dc average signal $I_0(\lambda)R(\lambda)$ and the ac modulated contribution $I_0(\lambda)\Delta R(\lambda)$, where $\Delta R(\lambda)$ is the modified reflectance resulting from the modulated perturbation. The complete signal feeds a lock-in amplifier (Stanford Instruments), which tracks the ac signal at the chopping frequency. The relative change in reflectance is obtained thereof by normalizing the ac signal with respect to the dc component, with typical values in the range of $10^{-3}$ to $10^{-6}$.

Figure 2 shows recorded spectra as a function of wavelength between 400 and 1100 nm under different pump beams and pass-filters. Long- (LPF) and short-pass filters (SPF) are indicated in the figure together with the nominal edge. The upper panel shows three measurements performed under 325 nm pump and different filter combinations: (i) LPF395 nm; (ii) LPF395 nm and LPF665 nm; and (iii) LPF395 nm and SPF600 nm. LPF395 nm prevents scattered laser light entering in the detector. Additional LPF665 nm and SPF600nm further restrict the accessible wavelength range towards higher or lower wavelengths from their nominal edge, respectively. Three PR signatures are readily observed in the figure, corresponding to $E_0$, $E_0+\Delta_0$, and $E_1$ transitions, as shown previously in Figure 1. Such interband transitions are well documented: $E_0$ corresponds to the lowest direct gap at the $\Gamma$ point of the Brioullin zone between $\Gamma_8$ valence- and $\Gamma_6$-conduction-band states; $E_0+\Delta_0$ corresponds to the split-off valence band $\Gamma_7$ due to spin-orbit coupling, connecting to the same $\Gamma_6$-conduction-band state; finally, $E_1$ is the next critical point in order of ascending energy and takes place along the $\Lambda$ direction from the center of the Brioullin zone [13]. The filter edges can be identified in the spectra with the declining signals deviating from the LPF395nm spectrum. Perfect overlapping over the respective wavelength ranges with the measurement using just LPF395 is observed, confirming the absence of eventual second-order harmonics in the spectra.

The medium panel shows spectra obtained under 632.8 nm pump illumination. The short wavelength spectrum was obtained with SPF600nm, whereas the long wavelength one was obtained with LPF665nm. The nominal wavelength of the laser is indicated by the dotted line. As it can be observed, the spectra collected under 632.8 nm pump keep track of $E_0$ and $E_1$ signatures ($E_0+\Delta_0$ is affected by the filter edges), very much like the 325 nm pump does, even when $E_1$ is not directly accessible now under 632.8 nm illumination. Instead, upstream modulation of high energy critical points results from absorption involving lower energy transitions $E_0$ and $E_0+\Delta_0$. The modified built-in potential and the associated field, due to photogenerated carrier screening at SCR, is the modulating mechanism affecting the entire electronic structure, including all high energy critical points. They can be probed thereof in a similar fashion as low energy critical points in downstream modulation. Finally, the lower panel of Figure 2 shows a PR spectrum obtained under 814 nm pump illumination using SPF800nm. The dotted line indicates the wavelength of the pump beam. Again, high energy critical points $E_1$ and $E_{SO}$ are readily probed when pumped with light of lower energy.

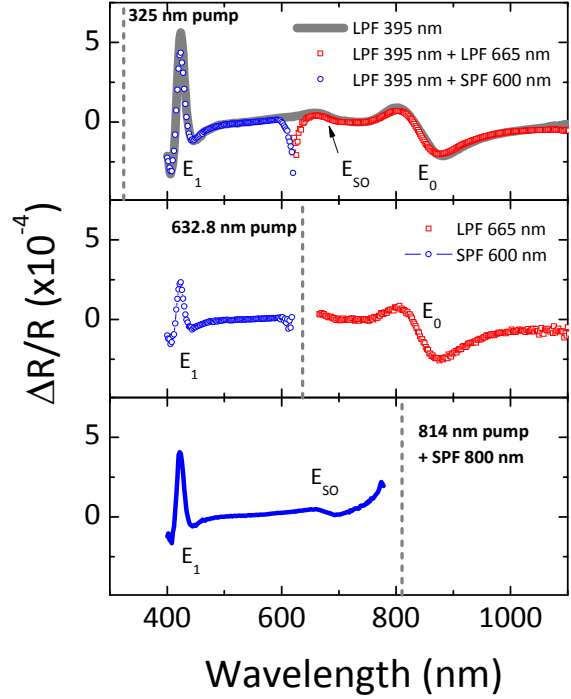

FIG. 2 PR spectra of *n*-GaAs wafer obtained under different pump beam energies and pass filters. Dashed lines indicate the nominal wavelength of the pump beams. Critical points $E_0$, $E_0+\Delta_0$ (labeled as $E_{SO}$), and $E_1$ are also indicated. (Upper panel) Using 325 nm pump with LPF395nm and additional LPF665nm or SPF600nm. (Middle panel) Using 632.8 nm pump with LPF665nm or SPF600nm. (Lower panel) Using 814 nm pump and SPF800nm.

Upstream photoreflectance is better understood when considering the character of modulation spectroscopies as absorption-based techniques. As such, and contrarily to the case of luminescence, PR also probes unoccupied states which are accessible to the energy range of the photons in the probe beam. However, it is not necessary that the pump generating the periodic perturbation be absorbed in a process involving that particular transition to be probed in the experiment. This result has been recently reported in GaSb [14] and previously in sub-bandgap PR on GaAs [15]. The latter case illustrates the fact that upstream modulation can also be activated via optically active defect states in the bandgap. As a matter of fact, the upstream energy range in PR has largely been overseen in the past, as evidenced by the absence of related literature, with just a few exceptions mentioned. Even in such cases, results have oftentimes been presented in relation to certain specificities of the samples, rather than as an expected output.

In summary, it has been shown that the information range accessible to PR can be extended to energies above that of the pump beam. Its practical implementation is simple, either replacing LPF with SPF or alternatively using notch or narrow-band filters around the wavelength of the pump beam. Probing upstream is a direct consequence of the absorption-based nature of the technique and the intrinsic modulation mechanism involved, based on photovoltage generation upon the action of the pump beam affecting the entire electronic structure of the material under test. Accounting for this fact, apparently not much explored yet, may widen the current applicability of the technique.

**Acknowledgements**

Financial support from the Ministry of Economy and Competitivity (projects TEC2015-64189-C3-1-R and AIC-B-2011-0806) and from Comunidad de Madrid (S2013/MAE-2780) is acknowledged.